\documentclass[11pt]{article}
\usepackage{xspace}
\usepackage{graphicx}
\usepackage{amsmath}
\usepackage{amssymb}
\usepackage{color}
\definecolor{Red}{rgb}{1,0,0}
\definecolor{Green}{rgb}{0,1,0}
\definecolor{Blue}{rgb}{0,0,1}
\definecolor{Black}{rgb}{0,0,0}

\def\beq{\begin{equation}}
\def\eeq#1{\label{#1}\end{equation}}
\def\eeqn{\end{equation}}

\def\beqa{\begin{eqnarray}}
\def\eeqa#1{\label{#1}\end{eqnarray}}
\def\eeqan{\end{eqnarray}}

\let\bar=\overbar

\def\etal{{\it et al.}}
\def\ie{{\it i.e.}}
\def\eg{{\it e.g.}}

\def\Dslash{\not{\hbox{\kern-4pt $D$}}}
\def\dslash{\not{\hbox{\kern-2pt $\del$}}}

\def\msb{{\bar{\ssstyle M \kern -1pt S}}}

\def\Title#1{\begin{center} {\Large {\bf #1} } \end{center}}
\begin{document}
\newcommand{\Kqz}{{\overline K}{}^0} 
\def\ri {{\rm i}} \def\re {{\rm e}} \def\rd {{\rm d}} \def\aq {{\overline a}{}}
\def\reps {{\rm Re}\,\epsilon} \def\rdel {{\rm Re}\,\delta} \def\Aq {{\overline A}{}}

\Title{T Violation in K-Meson Decays}
\bigskip\bigskip
\begin{raggedright}  
Klaus R.~Schubert\index{Schubert, K.~R.},\\ 
{\it Technische Universit\"at Dresden and
Johannes Gutenberg-Universit\"at Mainz}
\bigskip
\end{raggedright}
{\small
\begin{flushleft}
\emph{To appear in the proceedings of the 50 years of CP violation conference, 10 -- 11 July, 2014, 
held at Queen Mary University of London, UK.} \end{flushleft}}

\section{Introduction}

The first reported meson was a charged Kaon, observed with a cloud chamber in cosmic rays in 1944 
\cite{1944-LeprinceRinguet}, three years before the discovery of the charged pion \cite{1947-Lattes}. 
The neutral Kaon was also discovered in 1947 
\cite{1947-RochesterButler}. In 1955, Gell-Mann and Pais \cite{1955-GellMannPais} predicted that the 
$K^0$ is a two-state particle with a non-exponential decay law. This was confirmed in 1956 \cite{1956-Lederman} 
with 20 $K^0$ decays showing a life time at least 10 times longer than that of the dominant $K^0\to\pi^+\pi^-$ 
decays. In the same year, Lee and Yang \cite{1956-LeeYang} concluded that weak decays violate P symmetry since
the charged Kaon decays into $2\pi$ and $3\pi$ states with opposite parity. P violation was confirmed in two 
experiments \cite{1957-Wu, 1957-Lederman} one year later. In 1964, Chistenson \etal~ \cite{1964-CroninFitch} 
discovered that also CP symmetry is violated, either in decays of the long-living $K^0$ state 
$K_L$ (CP=-1) into $\pi^+\pi^-$ or in $K^0\Kqz$ transitions with mass eigenstates which are not CP eigenstates. 
A 1967 experiment \cite{1967-Steinberger} proved with
\beq \Delta_{Le} =\frac{N(K_L\to\pi^-e^+\nu)-N(K_L\to\pi^+e^-\nu)}{N(K_L\to\pi^-e^+\nu)+
     N(K_L\to\pi^+e^-\nu)}=(2.24\pm 0.36)~10^{-3} \label{Eq-1}\eeqn
that CP symmetry is violated in $K^0\Kqz$ transitions.

\section{Phenomenology of $K^0\Kqz$ Transitions}

Following Weisskopf and Wigner \cite{1930-WeisskopfWigner}, the evolution of the two-state neutral Kaon 
$|\Psi\rangle = \psi_1 |K^0\rangle + \psi_2 |\Kqz\rangle$ is given by the effective Schr\"odinger equation
\beq  
\ri~\frac{\partial}{\partial t }\left(\begin{array}{c}\psi_1\\ \psi_2\end{array}\right)=
{\cal H}_{eff} \left(\begin{array}{c}\psi_1\\ \psi_2\end{array}\right) =
\left[\left(\begin{array}{cc}m_{11}&m_{12}\\ m_{12}^*&m_{22}\end{array}\right)
-\frac{\ri}{2}\left(\begin{array}{cc}\Gamma_{11}&\Gamma_{12}\\
\Gamma_{12}^*&\Gamma_{22}\end{array}\right)\right]
\left(\begin{array}{c}\psi_1\\\psi_2\end{array}\right)~.\label{Eq-2}  
\eeqn
Owing to arbitrary phases of the states $|K^0\rangle$ and $|\Kqz\rangle$, the phases of $m_{12}$ and 
$\Gamma_{12}$ are unobservable. Their difference,
the phase of $\Gamma_{12}/m_{12}$, is an observable. In total, the equation 
has 7 real observable parameters:
$m_{11}$, $m_{22}$, $\Gamma_{11}$, $\Gamma_{22}$, $|m_{12}|$ and $|\Gamma_{12}|$ in 
addition to $\phi(\Gamma_{12}/m_{12})$. Two solutions
of Eq.~\ref{Eq-2} have exponential decay laws,
\beqa K^0_S(t) = [(1+\epsilon+\delta)\cdot K^0 +(1-\epsilon-\delta)\cdot\Kqz]\cdot 
\re^{-\ri m_S t-\Gamma_S t/2}/\sqrt{2}~,\nonumber\\
     K^0_L(t) = [(1+\epsilon-\delta)\cdot K^0 -(1-\epsilon+\delta)\cdot\Kqz]\cdot 
\re^{-\ri m_L t-\Gamma_L t/2}/\sqrt{2}~. \label{Eq-3}  \eeqan
They are normalized to 1 at $t=0$ and they are in general not orthogonal,
\beq \langle K_S|K_L\rangle = 2~{\rm Re}~\epsilon - 2~ \ri~ {\rm Im}~\delta~.\label{Eq-3b}\eeqn
The 7 observable parameters of the solutions, following
unambiguously from the 7 parameters in Eq.~\ref{Eq-2}, are $m_S$, $m_L$, $\Gamma_S$, $\Gamma_L$, Re\,$\epsilon$, 
Re\,$\delta$ and
Im\,$\delta$. The additional parameter Im\,$\epsilon$ is unobservable, and Eqs.~\ref{Eq-3} are approximations in 
the limits $|{\rm Re}~\epsilon|\ll 1$
and $|\delta|\ll 1$, well fulfilled experimentally. The relations between the parameter sets in Eqs.~\ref{Eq-3} and 
\ref{Eq-2} are well
approximated by
\beqa m_L=m+\Delta m/2~,~~m_S=m-\Delta m/2~&,&~~ \Gamma_L=m+\Delta \Gamma/2~,~~\Gamma_S=m-\Delta \Gamma/2~,\nonumber\\
      m =(m_{11}+m_{22})/2~&,&~~ \Gamma =(\Gamma_{11}+\Gamma_{22})/2~,\nonumber\\
      \Delta m = 2 |m_{12}|~&,&~~\Delta\Gamma = 2 |\Gamma_{12}|\times {\rm sign[Re}(\Gamma_{12}/m_{12})]~,\nonumber\\
      {\rm Re}~\epsilon = \frac{{\rm Im}(\Gamma_{12}/m_{12})}{4+|\Gamma_{12}/m_{12}|^2}~&,&~~\delta = \frac{(m_{22}-m_{11})
      -\ri(\Gamma_{22}-\Gamma_{11})/2}{2\Delta m-\ri\Delta\Gamma}~.\label{Eq-4}  \eeqan
The choice $\Delta m > 0$ is a convention, the sign of $\Delta\Gamma$ is given by $\Gamma_{12}/m_{12}$. CPT symmetry 
requires $\delta=0$, T symmetry
Re~$\epsilon =0$, \ie~$\phi(\Gamma_{12}/m_{12}) =0$ or $\pi$, and CP symmetry Re~$\epsilon =\delta =0$.  
T symmetry requiring Im~$(\Gamma_{12}/m_{12}) =0$ has the same origin as requiring 
Im~$(A_{\rm M1}/A_{\rm E2})=0$ in atomic transitions with vanishing E1 amplitude, as shown by Lloyd in 1951 \cite{1951-Lloyd}.

\section{Motion-Reversal Symmetry in Classical Mechanics}

\begin{figure}[h]
\begin{minipage}[t]{0.30\textwidth}
\includegraphics[width=\textwidth]{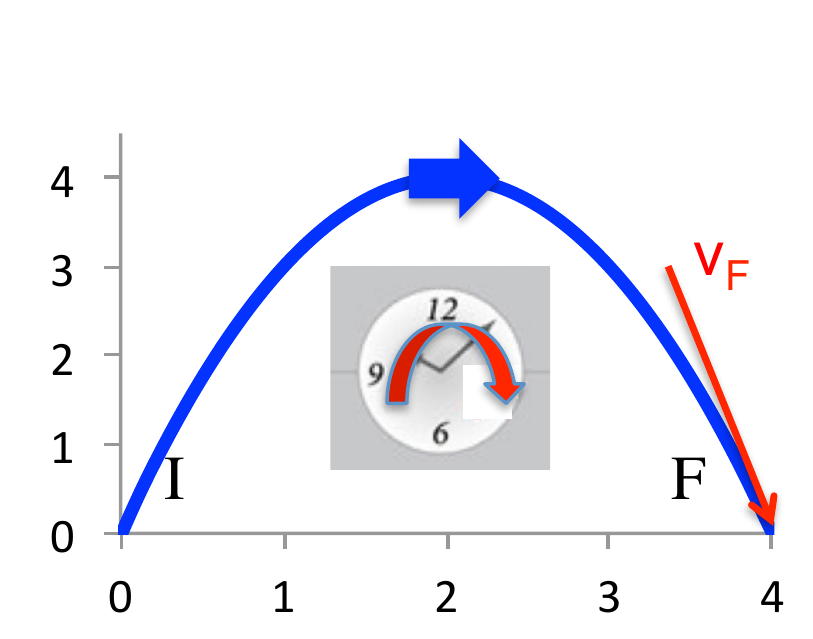}
\caption{Video recording of a motion with velocity ${\vec v}{}_F$ in the endpoint F together with 
a clock showing the running time.}
\label{Fig-2a} 
\end{minipage}\hfill
\begin{minipage}[t]{0.30\textwidth}
\includegraphics[width=\textwidth]{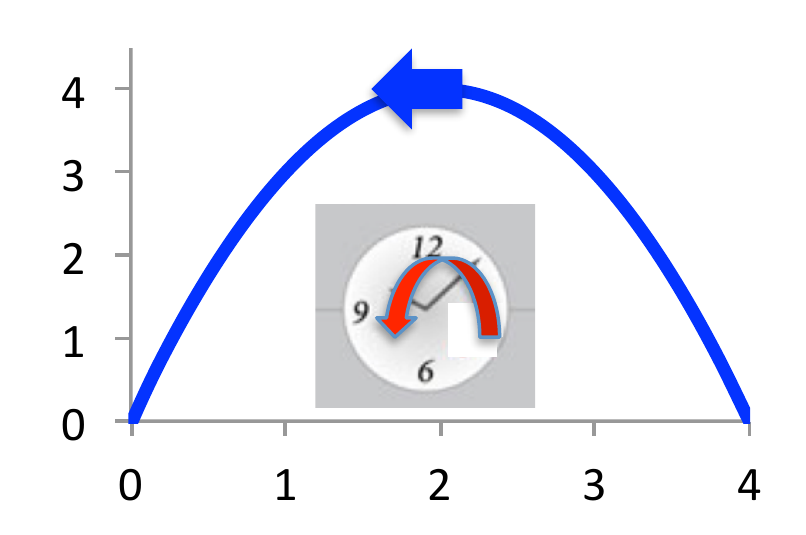}
\caption{Playback of the recorded video in the reversed time-direction.}
\label{Fig-2b} 
\end{minipage}\hfill
\begin{minipage}[t]{0.30\textwidth}
\includegraphics[width=\textwidth]{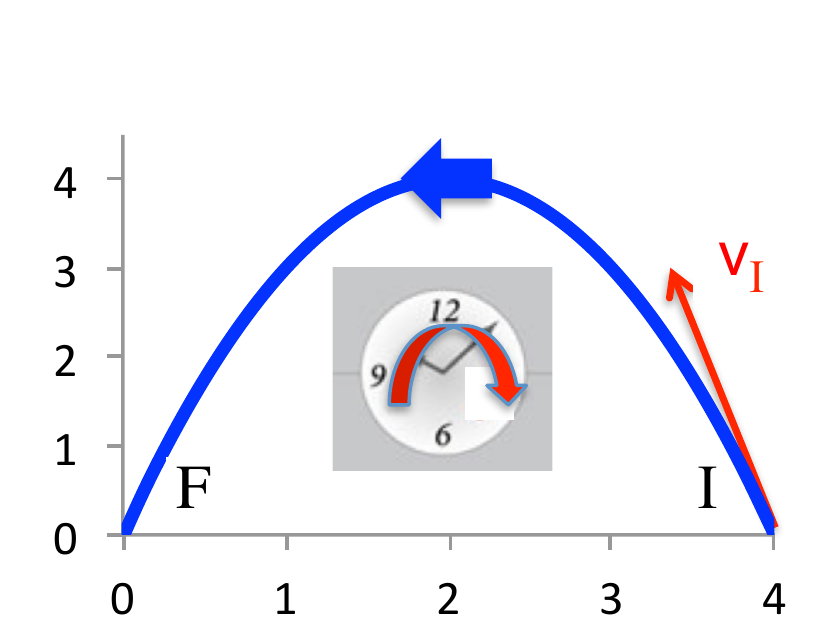}
\caption{The time-reversed motion with velocity ${\vec v}{}_I=-{\vec v}{}_F$ in the new start 
point I together with a clock
showing the forward-running time.}
\label{Fig-2c} 
\end{minipage}
\end{figure}

Fig.~\ref{Fig-2a} shows the orbit of a ball with small velocity on the surface of the earth. 
The motion can be recorded together with a clock showing the time from the start to the end of the orbit.
A video recorder allows to replay the recorded movie in the backward direction, as shown in Fig.~\ref{Fig-2b}.
Even if the motion of the ball looks familiar to the viewer, the clock tells him that this process does not take place
in his world. The replayed movie shows an unobservable process; time never runs backwards. However, it is possible
to observe the reversed motion in the real world: we have to start it at the end of the original orbit with a
velocity vector equal to the opposite of the final velocity in the original motion, as shown in Fig.~\ref{Fig-2c}.
This observable motion with forward-running time is called ``reversed motion". The operation with starting at the
end point and reversing the velocity is called ``motion reversal", and the comparison of Figs.~\ref{Fig-2c} and \ref{Fig-2a}
shows ``motion-reversal symmetry".

The T operation, time reversal, is defined by the transformation $t\to -t$ in the equation of motion, here
\beq m~¸\rd^2 {\vec r}/\rd t^2 = m~{\vec g}~. \eeqn
This equation is invariant if $t$ is replaced by $-t$, the equation is T-symmetric. Motion-reversal symmetry is a
consequence of T symmetry. Friction in the air leads to an equation of motion such as
\beq m~¸\rd^2 {\vec r}/\rd t^2 = m~{\vec g}-\eta~\rd{\vec r}/\rd t, \label{Eq-5}\eeqn
which is not invariant under the T transformation if $\eta\ne 0$. Consequently, motion-reversal symmetry is violated,
known since
many centuries from the asymmetric orbits of cannon balls. There are two ways to test T symmetry: either by
``direct" observation of motion-reversal violation (Move the cannon and fire from the target position!) or by
``indirect" observation of T violation (Measure the parameter $\eta$ in Eq,~\ref{Eq-5}!). 

\section{Time Reversal in Quantum Mechanics}

The T operation in quantum mechanics was introduced by Wigner in 1932 \cite{1932-Wigner}. Here follows a short summary
of the concept and its implications: Together with the commutation relation $[x_i,p_j]=\ri~\delta_{ij}$, the operation
$t\to -t$ leads to
\beq T~\ri~ T^{-1}=-\ri~,~~T=U~K~,\label{Eq-6}\eeqn
where $U$ is an arbitrary unitary transformation, $U~U^\dagger=1$, and $K$ is complex conjugation, $K~z~K^{-1}=z^*$,
leading to
\beq T~|\psi\rangle = U~|\psi^*\rangle~.\label{Eq-7}\eeqn
$T$ is antiunitary, $T^\dagger \ne T^{-1}$, and antilinear, $T(c_1|\psi_1\rangle+c_2|\psi_2\rangle)=
c_1^*|\psi_1^*\rangle + c_2^*|\psi_2^*\rangle$, with
\beq \langle \psi_{1T}|\psi_{2T}\rangle = \langle \psi_{2}|\psi_{1}\rangle~.\label{Eq-8}\eeqn
The dynamics of transitions and decays is described by operators $\cal H$, $S$ and $D$ with
\beq S = {\rm lim}~\re^{-\ri{\cal H}t} = 1 + \ri D~,~~S S^\dagger =1~,~~TDT^{-1}=D^\dagger~.\label{Eq-9}\eeqn
States of one or more particles with momenta $p_i$ and spins $s_i$ transform like
\beq T~|p_i, s_i\rangle = \re^{\ri\phi}~|-p_i, -s_i\rangle~\label{Eq-10}\eeqn
with an arbitrary phase $\phi$. This leads to
\beq T~\langle p_f, s_f|D|p_i, s_i\rangle = \re^{\ri(\phi_i-\phi_f)}~\langle-p_i, -s_i|D|-p_f, -s_f\rangle ^*~,
\label{Eq-12}\eeqn
and T symmetry requires ``motion-reversal symmetry",
\beq |\langle p_f, s_f|D|p_i, s_i\rangle|^2 = |\langle-p_i, -s_i|D|-p_f, -s_f\rangle|^2~.\label{Eq-13}\eeqn
When motion-reversal symmetry is observed to be violated, \eg ~in the $K^0$ system with $p=s=0$,
\beq |\langle \Kqz|D|K^0\rangle|^2 \ne |\langle K^0|D|\Kqz\rangle|^2~,\label{Eq-14}\eeqn
then the T symmetry of $\cal H$, $S$ and $D$ in the dynamics of $K^0 \Kqz$ transitions is violated, \ie ~Re~
$\epsilon\ne 0$.

I would like to add an important side remark: T violation in the Hamiltonian of an interaction is different from
 the omnipresent ``arrow of time". The entropy increase of an ensemble of $10^8$ $K^0$ mesons in $10^{-8}$ s is
 huge and indistinguishably equal for the two cases without or with a small T violation. 

\section{CPT}

The CPT operation is defined as $CPT = CP \times T$, and for bosonic systems we have $T^2 = +1$; therefore 
\beq {\Large CP = CPT \times T~}.\eeqn
When CP symmetry is broken, CPT or T or both must also be broken. The case with CP conservation 
and violation of both CPT and T is not possible in $K^0\Kqz$ transitions since its dynamics, 
see Eq.~\ref{Eq-2}, contains only 7 parameters. The 1967 CP violation in Eq.~\ref{Eq-1} can have two contributions, 
T violation with Re~$\epsilon\ne 0$
and CPT violation with $\delta\ne 0$. Two experimental ways have been used to determine the two contributions:
\begin{itemize}
\item{``Direct" measurements of T violation by motion reversal comparing\\
$|\langle K^0|U(t)|\Kqz\rangle|^2$ and $|\langle \Kqz|U(t)|K^0\rangle|^2$, and of CPT violation by comparing\\
$|\langle K^0|U(t)|K^0\rangle|^2$  and
$|\langle \Kqz|U(t)|\Kqz\rangle|^2$, where $U(t)=\re^{-\ri{\cal H}t}$},
\item{``Indirect" determinations of Re~$\epsilon$ and $\delta$ using Bell-Steinberger's unitarity relation.}
\end{itemize}
The direct way has been used by CPLEAR in 1998 \cite{1998-CPLEAR-T, 1998-CPLEAR-CPT}, the indirect way in 1970 by
Schubert \etal~\cite{1970-Schubert}. Both lead to the same results as discussed in the following sections.

\section{Unitarity Relations}

For an unstable single-state particle with the wave function $\psi(t)=\re^{-\Gamma t/2}$, unitarity
(conservation of probability including all observable decay states $f_i$) requires
\beq |\psi(t)|^2+\sum_1^N {|f_i(t)|^2} = 1~,~~\rd|\psi|^2/\rd t + \sum_1^N {\rd|f_i(t)|^2/\rd t}=0~. \label{Eq-20}\eeqn
At $t=0$, we have $\rd|\psi|^2/\rd t=-\Gamma$ and $\rd|f_i(t)|^2/\rd t=|\langle f_i|D|\psi\rangle|^2$ and 
\beq \Gamma = \sum_1^N {|\langle f_i|D|\psi\rangle|^2}~.\label{Eq-20b}\eeqn
This unitarity relation connects the sum of all decay rates to the inverse mean life of the unstable particle.

For the two-state particle $\Psi(0) = \psi_1 K^0 + \psi_2 \Kqz = \alpha K_S + \beta K_L$, we obtain
three unitarity relations. As presented \eg~in Ref.~\cite{BrancoLavouraSilva}, unitarity in the space of states
$K_S, K_L, f_1 \dots f_N$ leads to the three relations
$$ \Gamma_S =  \sum_1^N|\langle f_i|D|K^0_S\rangle|^2~,~~\Gamma_L =\sum_1^N|\langle f_i|D|K^0_L\rangle|^2~,$$
\beq\left(\frac{\Gamma_S+\Gamma_L}{2}+\ri\Delta m\right)\langle K^0_S|K^0_L\rangle~=~\sum_1^N
     \langle f_i|D|K^0_S\rangle^* \langle f_i|D|K^0_L\rangle ~.\label{Eq-21}\eeqn
Using the result for $\langle K^0_S|K^0_L\rangle$ in Eq.~\ref{Eq-2}, the third relation can be written as
\beq {\rm Re}\,\epsilon-\ri~{\rm Im}\,\delta =\frac {\sum_1^N\langle f_i|D|K^0_S\rangle^* \langle f_i|D|K^0_L\rangle }
    {\Gamma_S+\Gamma_L+2\ri\Delta m}~.\label{Eq-22}\eeqn
It was derived in 1966 by Bell and Steinberger \cite{1966-BellSteinberger}
and allows to determine the two $K^0\Kqz$ transition parameters ${\rm Re}\,\epsilon$ and ${\rm Im}\,\delta$ 
using measurable decay properties, as described in the following. 

\begin{figure}[h]
\begin{minipage}{0.32\textwidth}
\includegraphics[width=\textwidth]{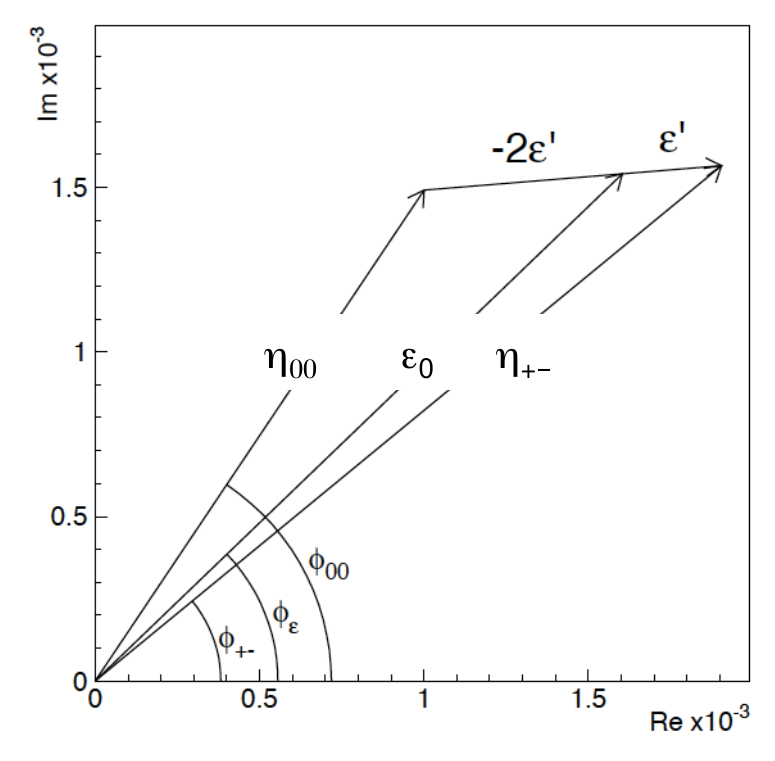}
\caption{The Wu-Yang triangle \cite{1964-WuYang} relating the parameters $\eta_{+-}$, $\eta_{00}$, 
$\epsilon_W$ and $\epsilon^\prime$.} 
\label{Fig-WuYang} 
\end{minipage}\hfill
\begin{minipage}{0.64\textwidth}
The dominant final states are $\pi^+\pi^-$ and $\pi^0\pi^0$ with the well-known CP-violation parameters
\beq \eta_{+-} = \frac {\langle \pi^+\pi^-|D|K^0_L\rangle}{\langle \pi^+\pi^-|D|K^0_S\rangle}=\epsilon_W
+\epsilon^\prime~,\label{Eq-24}\eeqn
\beq  \eta_{00} = \frac {\langle \pi^0\pi^0|D|K^0_L\rangle}{\langle \pi^0\pi^0|D|K^0_S\rangle}=\epsilon_W
-2~\epsilon^\prime~.\label{Eq-25}\eeqn
Note that the $\eta$ parameter for $\pi\pi$ with isospin 0 is called $\epsilon_W$ here (W for Wolfenstein) in order
to distinguish it from the $K^0\Kqz$ transition parameter $\epsilon$.  The $\eta$ parameter $\epsilon_W$ has an
observable phase, $\epsilon$ has not. In the first four years after the discovery of CP violation, $|\eta_{+-}|$,
$\phi(\eta_{+-})$ and $|\eta_{00}|$ were determined precisely enough for being used in a Bell-Steinberger unitarity
analysis, but there was no result on $\phi(\eta_{00})$ before 1970.
\end{minipage}\end{figure}

{\noindent The Bell-Steinberger relation may be written as}
\beq {\rm Re}\,\epsilon-\ri~{\rm Im}\,\delta =\frac {\epsilon_W+\sum_{i\ne\pi\pi}\alpha_i}{1+\Gamma_L/\Gamma_S+2~\ri~
    \Delta m/\Gamma_S}~{\rm with}~\alpha_i = \frac{\langle f_i|D|K^0_L\rangle}{\langle f_i|D|K^0_S\rangle}\times BF(K_S\to f_i)
    ~,\label{Eq-26}\eeqn
where $BF$ is the branching fraction. Since $\Gamma_L/\Gamma_S\ll 1$, the phase $\phi(\epsilon_W)$ of 
$\epsilon_W=(2~\eta_{+-}+\eta_{00})/3$ shows the two contributions
of CP violation in $K^0\Kqz$ transitions if $|\sum \alpha_i|\ll |\epsilon_W|$:
\begin{itemize}
\item{CP violation with CPT symmetry, $\delta=0,~ \Rightarrow~ \phi(\epsilon_W)=\arctan(\Delta m/\Gamma_S)\approx 45^\circ$~,}
\item{CP violation with T symmetry, Re~$\epsilon=0,~ \Rightarrow~\phi(\epsilon_W)=\arctan(-\Delta m/\Gamma_S)\approx 135^\circ$~.}
\end{itemize}
If $\phi(\epsilon_W)$ is neither $45^\circ$ nor $135^\circ$, both T and CPT violation contribute. In 1968, only $\phi(\eta_{00})$ 
was missing for a determination of the two contributions, if $\sum\alpha_i$ for $i\ne\pi\pi$ is well enough estimated.

\section{First Measurement of $\phi(\eta_{00})$ and First Observation of T Violation}

\begin{figure}[h]
\begin{minipage}{0.40\textwidth}
\includegraphics[width=\textwidth]{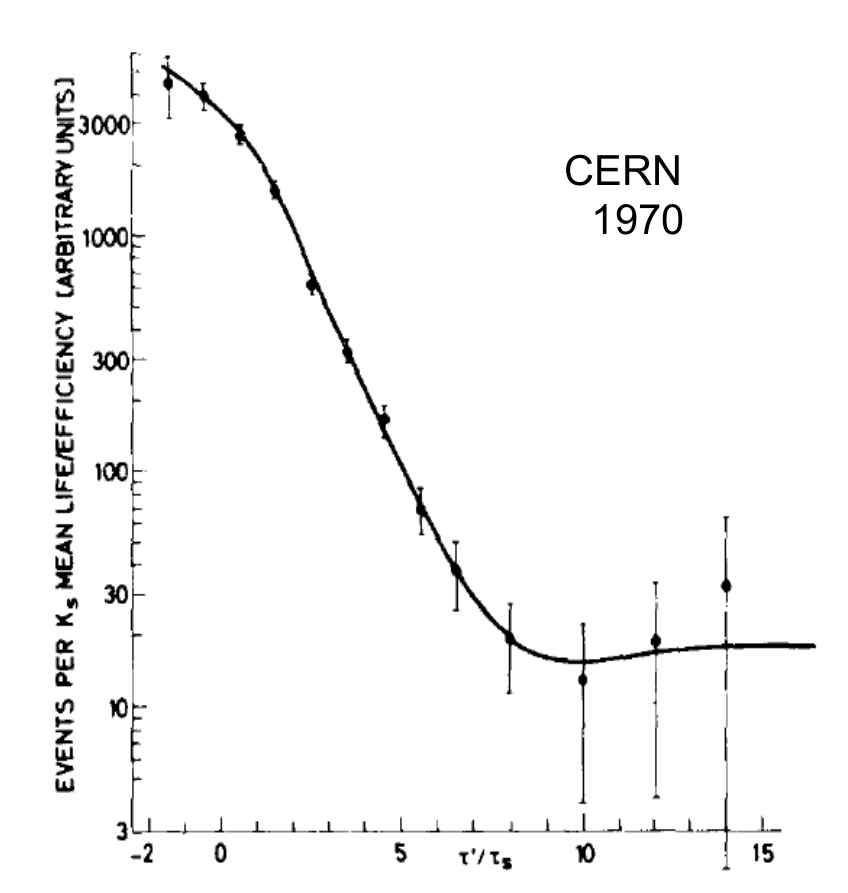}
\caption{Lifetime distributions of $\pi^0\pi^0$ events from neutral Kaons behind a Cu regenerator \cite{1970-Chollet}.} 
\label{Fig-Chollet} 
\end{minipage}\hfill
\begin{minipage}{0.55\textwidth}
These two results have been used in the same year by the same group \cite{1970-Schubert} for the first Bell-Steinberger
analysis together with the results $2\Delta m/\Gamma_S=0.983\pm 0.0030)$, $|\eta_{+-}|=(1.92\pm 0.05)\times 10^{-3}$,
$\phi(\eta_{+-})=(44\pm 5)^\circ$,
$\alpha_{3\pi, I=1} = [0.3\pm 1.7 +\ri (0.8\pm 2.5)]\times 10^{-4}$, 
$\alpha_{\pi e\nu + \pi\mu\nu} = [-2.4\pm 2.1 +\ri (0.9\pm 3.4)]\times 10^{-4}$, and well-moitivated zero $\alpha$ values for
$3\pi, I=3$, $\pi\pi\gamma$ and $\gamma\gamma$. The analysis results in
\beqa {\rm Re}~\epsilon &=& (1.68\pm 0.30)\times 10^{-3}~,\nonumber\\
{\rm Im}~\delta &=& (-0.30\pm 0.45)\times 10^{-3}~,\label{Eq-28} \eeqan
T violation in $K^0\Kqz$ transitions is established with 5 $\sigma$, and CPT symmetry is found to be valid within errors. 
The analysis cannot determine Re~$\delta$, only the CPT-violating quantity
\beq {\rm Re}~(\delta+{\tilde\alpha}{}_0) = (0.07\pm 0.43)\times 10^{-3}~,\label{Eq-29}\eeqn
\end{minipage}\end{figure}

The group of J.~M.~Gaillard at CERN \cite{1970-Chollet} determined in 1970 the time dependence of the decay rate
$|\langle\pi^0\pi^0|D|\Psi(t)\rangle|^2$ after a copper regenerator with $\psi(0)=K_L+\rho~K_S$ using $\sim 200~000$ photographs
from a setup of optical spark chambers with a scintillation-counter trigger. The result is shown in Fig.~\ref{Fig-Chollet}, and the
best fit to the data gives 
\beq |\eta_{00}|=(3.3\pm 0.7)\times 10^{-3}~,~~\phi(\eta_{00})=(51\pm 30)^\circ~.\label{Eq-27}\eeqn

{\noindent where ${\tilde\alpha}{}_0$ describes CPT violation in $K^0\to\pi\pi_{I=0}$ decays.}
 
\section {Updates of Bell-Steinberger Analyses}

The following list of updates is far from being complete. It gives some milestones, each time driven by more precise data.
\newpage
\begin{table}[!th]
\begin{center}
\caption{Selected milestones of Bell-Steinberger analyses. Comments are given below the Table.
For the Re~$\delta$ results, see Section \ref{Sec-PiEllNu}.}  
\begin{tabular}{|l|l|c|c|c|} \hline\hline
Year & Reference & Re~$\epsilon~[10^{-3}]$ & Im~$\delta~[10^{-3}]$ & Re~$\delta~[10^{-3}]$\\ \hline
1970 & Schubert \etal~\cite{1970-Schubert} & $ 1.68\pm 0.30 $  & $-0.30\pm 0.45$ & \\
1980 & Cronin \cite{1980-Cronin} & $ 1.61\pm 0.20 $  & $-0.08\pm 0.17$  & \\
1983 & ITEP Moscow \cite{1983-BaldoCeolin} & $ 1.62\pm 0.05 $  & $-0.11\pm 0.10$  & \\
1999 & CPLEAR \cite{1999-CPLEAR-BellSt} & $ 1.649\pm 0.025 $  & $-0.02\pm 0.05$  & $0.24\pm 0.27$\\
2006 & KLOE \cite{2006-KLOE-BellSt} & $ 1.596\pm 0.013 $  & $0.004\pm 0.021$  & $0.23\pm 0.27$ \\
2012 & Particle Data Group \cite{2012-PDG, 2012-PDG-BellSt} & $ 1.611\pm 0.005 $  & $-0.007\pm 0.014$  & $0.24\pm 0.23$\\
\hline\hline \end{tabular} \label{Tab-BellSteinberger} \end{center} 
\end{table}

The main progress in the 1980 analysis originates from the new result $\phi(\eta_{00})=(56\pm 6)^\circ$ obtained by Christenson 
\etal~\cite{1979-Christenson}. The 1983 analysis profits from new data on $\pi^+\pi^-\pi^0$ decays and on the first determination
of $\eta_{000}=\langle\pi^0\pi^0\pi^0|D|K_S\rangle/\langle\pi^0\pi^0\pi^0|D|K_L\rangle=(-0.08\pm 0.18)+\ri(-0.05\pm 0.27)$ by
Barmin \etal~\cite{1983-Barmin}. The 1999 analysis uses the rich CPLEAR data set of $10^8~K^0$ decays into $\pi^+\pi^-$, 
$\pi^0\pi^0$, $\pi e\nu$, $\pi^+\pi^-\pi^0$ and $3\,\pi^0$. Its first result on Re~$\delta$ is obtained from the determination
of all $\pi e \nu$ decay parameters in combination with Bell-Steinberger unitarity \cite{1999-CPLEAR-BellSt}. The 2006 analyses
uses the even larger KLOE data set with $2.5\times 10^9~e^+e^-\to\Phi\to K^0\Kqz$ events with neutral-Kaon decays into
essentially all final states. 

\begin{figure}[h]
\begin{minipage}{0.40\textwidth}
\includegraphics[width=\textwidth]{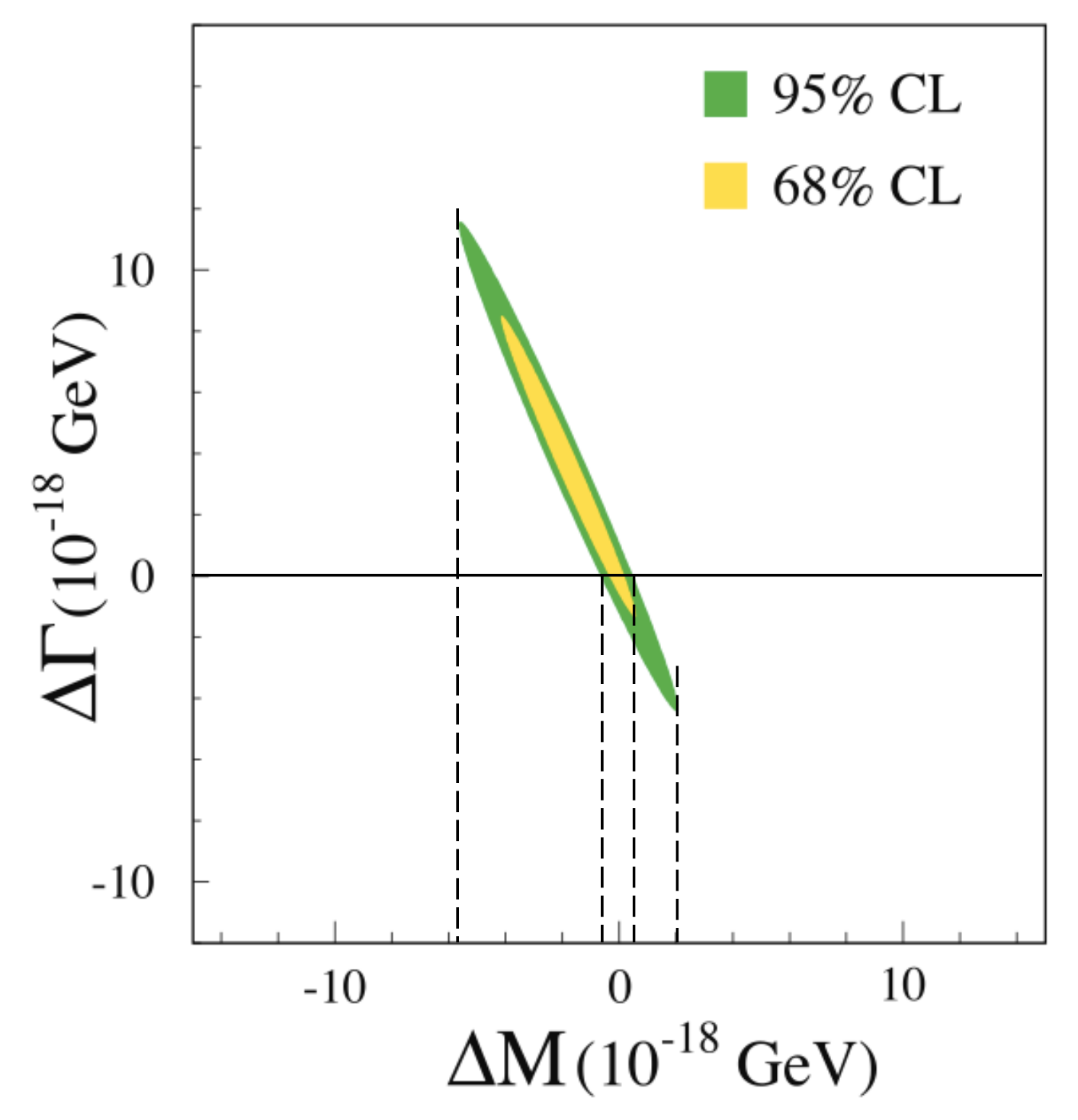}
\caption{Likelihood contours for $-\delta m$ vs.~$-\delta\Gamma$ \cite{2012-PDG-BellSt}. The outer two and inner two dashed lines 
correspond to the two cases in Eqs.~\ref{Eq-30}.} \label{Fig-delMdelG}
\end{minipage}\hfill
\begin{minipage}{0.55\textwidth}
The 2012 analysis of Antonelli and D'Ambrosio uses the latest PDG averages for all inputs
including NA48 and KTeV results on $2\pi$ decays to be discussed in Section \ref{Sec-PiPi}.
With the definition of $\delta$ in Eq.~\ref{Eq-4}, the obtained values for real and imaginary part of $\delta$ determine the
CPT-violating parameters in the Schr\"odinger equation Eq.~\ref{Eq-2},
\beqa \delta m = m_{22}-m_{11} = m(\Kqz)-m(K^0)~,\nonumber\\~~\delta\Gamma = \Gamma_{22}-\Gamma_{11} = \Gamma(\Kqz)-\Gamma(K^0)~.\eeqan
The results are shown in Fig.~\ref{Fig-delMdelG}. For the mass difference, two different results may be given, with no constraint on 
$\delta\Gamma$ (first line) and with $\delta\Gamma=0$ (second line):
\beqa -2\times 10^{-18} < \delta m < +6\times 10^{-18}~{\rm GeV},\nonumber\\
      -4\times 10^{-19} < \delta m < +4\times 10^{-19}~{\rm GeV}.\label{Eq-30}\eeqan
Note that the assumption $\delta\Gamma = 0$ also implies Re~$\delta\approx$ Im~$\delta$, since $\Delta m\approx \Delta\Gamma/2$.
\end{minipage}\end{figure}

\section {$\pi\ell\nu$ Decay Amplitudes} \label{Sec-PiEllNu}

The four decay amplitudes, all understood to be integrals over the three-particle phase spaces,
\beq \langle\pi^-\ell^+\nu|D|K^0\rangle~,~~\langle\pi^+\ell^-\nu|D|\Kqz\rangle~,~~
     \langle\pi^-\ell^+\nu|D|\Kqz\rangle~,~~\langle\pi^+\ell^-\nu|D|K^0\rangle~,\label{Eq-31}\eeqn
define six real parameters $a$, $y$, Re~$x_+$, Im~$x_+$, Re~$x_-$ and Im~$x_-$. When $a$ is much larger
than the other five parameters, which is well fulfilled experimentally, they are defined as
\beqa a=\frac{\langle\pi^-\ell^+\nu|D|K^0\rangle + \langle\pi^+\ell^-\nu|D|\Kqz\rangle}{2},~
      y=\frac{\langle\pi^+\ell^-\nu|D|\Kqz\rangle-\langle\pi^-\ell^+\nu|D|K^0\rangle}{2~a}~,\nonumber\\
      x_+=\frac{\langle\pi^-\ell^+\nu|D|\Kqz\rangle + \langle\pi^+\ell^-\nu|D|K^0\rangle^*}{2~a},~
      x_-=\frac{\langle\pi^-\ell^+\nu|D|\Kqz\rangle - \langle\pi^+\ell^-\nu|D|K^0\rangle^*}{2~a}.\label{Eq-32}\eeqan
The parameters $a$ and $y$ obey the so-called ``$\Delta Q=\Delta S$" rule, the two $x$ parameters violate it,
$a$ obeys and $y$ violates CPT. $x_+$ obeys and $x_-$ violates CPT. The real parts of $x_+$ and $x_-$ obey and the
two imaginary parts violate T. Im~$x_-$ is a very special quantity. If non-zero, it violates all three symmetries
CP, T and CPT.

Measuring the time dependences of the four decay rates in Eq.~\ref{Eq-31}, CPLEAR obtained in 
1998 from $1.3\times 10^6~\pi e\nu$ events
\beqa {\rm Re}~x_+= (-1.8\pm 6.1)\times 10^{-3}~,~~{\rm Im}~x_+= (1.2\pm 2.1)\times 10^{-3}~,\nonumber\\
      {\rm Im}~x_-= (-0.8\pm 3.5)\times 10^{-3}~,~~{\rm Re}~\delta = (0.30\pm 0.34)\times 10^{-3}~,\eeqan
where the values are taken from the 2003 CPLEAR summary report \cite{2003-CPLEAR-report} and the quoted errors 
combine statistics and systematics.

When the time dependences are combined with Bell-Steinberger unitarity \cite{1999-CPLEAR-BellSt} and with the 1999 world average for
the asymmetry $\Delta_{Le}$ as already defined in Eq.~\ref{Eq-1}, the CPLEAR fits improve
the sensitivity on Re~$\delta$ and become sensitive to the CPT-violating parameters $y$ and Re~$x_-$ independent of any
additional assumptions,
\beq {\rm Re}~\delta = (0.24\pm 0.27)\times 10^{-3},~y=(0.3\pm 3.1)\times 10^{-3},~{\rm Re}~x_-= (-0.5\pm 3.0)\times 10^{-3}.
      \label{Eq-34}\eeqn
The error on Re~$\delta$ improves, no CPT violation is seen in $K^0\Kqz$ transitions. In addition, no violation of the
``$\Delta Q=\Delta S$" rule is seen, the two $x$ parameters are compatible with zero. Therefore, there is no visible T violation in the
$\pi\ell\nu$ decay amplitudes. Any CP violation therein could only come from CPT violation. Since $y$ is also compatible with zero,
the decay amplitudes are symmetric under CPT, T and CP. 

\section {Transverse Muon Polarisation in $K^0\to\pi\mu\nu$ Decays} 

T symmetry of the Hamiltonian led to the result in Eq.~\ref{Eq-12},
\beq \langle p_f, s_f|D|p_i, s_i\rangle = \re^{\ri(\phi_i-\phi_f)}~\langle-p_f, -s_f|D^\dagger|-p_i, -s_i\rangle ~. \eeqn
If decays are only influenced by weak interactions (Standard and weaker), \ie~if stronger final-state interactions (FSI) are
absent, unitarity of the $S$ operator, $S=1+\ri~D,~S S^\dagger =1$ leads to
\beq D-D^\dagger=\ri~D^\dagger D\approx 0~,~~D^\dagger \approx D~,\eeqn
since the second-order interaction $D^\dagger D$ is much smaller than $D$. Consequently,
\beq \langle p_f, s_f|D|p_i, s_i\rangle \approx \re^{\ri(\phi_i-\phi_f)}~\langle -p_f, -s_f|D|-p_i, -s_i\rangle ~,\eeqn
which is called $\hat{T}$ symmetry in the textbook of Branco, Lavoura and Silva \cite{BrancoLavouraSilva}.
For decays of $K$ mesons in their rest frame this means in very good approximation
\beq |\langle p_f, s_f|D|0,0\rangle| = |\langle-p_f, -s_f|D|0,0\rangle| ~. \eeqn
For the polarization triple product in decays $K^0_L\to\pi^-\mu^+\nu$ and $K^+\to\pi^0\mu^+\nu$ this requires
\beq {\vec s}\cdot [{\vec p_\pi} \times{\vec p_\mu} ]= -{\vec s}\cdot [{\vec p_\pi} \times{\vec p_\mu} ]=0\eeqn
at any point $(|{\vec p}_\mu|,~\theta_{\mu\nu})$ of the Dalitz plot. This can be parametrised as
\beq \frac{{\vec s}\cdot [{\vec p_\pi} \times{\vec p_\mu} ]}{|{\vec p_\pi}\times{\vec p_\mu}|}={\rm Im}~\xi~\frac{m_\mu}{m_K}~
    f(|{\vec p}_\mu|,~\theta_{\mu\nu})~.\eeqn
Since there is only one hadron in the final state, no strong interaction can mimic T violation, and electromagnetic FSI are
estimated to produce Im~$\xi\approx 0.008$ \cite{1973-GinsbergSmith}.
A BNL experiment in 1980 \cite{1980-Morse} finds Im~$\xi = 0.009\pm 0.030$ for $K^0_L\to\pi^-\mu^+\nu$, and a 2006 KEK experiment
\cite{2006-Abe} finds Im~$\xi = -0.005\pm 0.008$ for $K^+\to\pi^0\mu^+\nu$. There is no T violation within one standard deviation.
Both experiments test T violation in New Physics, Standard Model estimates expect Im~$\xi\approx 10^{-7}$ without FSI 
\cite{2000-BigiSanda}. 
   
\section {``Direct" Tests of T and CPT Violation in $K^0\Kqz$ Transitions} \label{Sec-DirectTandCPT}

The explicit time dependences for the appearance of $\Kqz$ ($K^0$) states from initial $K^0$ ($\Kqz$) states can easily be 
derived from Eqs.~\ref{Eq-3}; they are found to be
\beqa P(K^0\to\Kqz) =  (\frac{1}{4} - {\rm Re}~\epsilon)~
                       (\re^{-\Gamma_S t}+\re^{-\Gamma_L t}-2\cos\Delta mt\cdot\re^{-\Gamma t}) ~,\nonumber\\
     P(\Kqz\to K^0) =  (\frac{1}{4} + {\rm Re}~\epsilon)~
                       (\re^{-\Gamma_S t}+\re^{-\Gamma_L t}-2\cos\Delta mt\cdot\re^{-\Gamma t})~. \eeqan
They result in a motion-reversal asymmetry
\beq A_{T}(t) = \frac {P(\Kqz\to K^0)-P(K^0\to\Kqz)}{P(\Kqz\to K^0)+P(K^0\to\Kqz)} = 4~{\rm Re}~\epsilon~,\label{Eq-40}\eeqn
which is time-independent and given by the only T-violating parameter in $K^0\Kqz$ transitions, Re~$\epsilon$. 
Observation of this asymmetry requires preparation of the initial states and detection of the final states. CPLEAR 
\cite{1998-CPLEAR-T} prepares
the initial states by the reactions $p{\overline p}\to K^0 K^-\pi^+,~\Kqz K^+\pi^-$, detects the final states by 
decays into $\pi^- e^+\nu$ and $\pi^+ e^-\nu$ and determines the asymmetry
\beq A_T^{exp}(t) = \frac {P(\Kqz\to \pi^- e^+\nu)-P(K^0\to\pi^+ e^-\nu)}{P(\Kqz\to \pi^- e^+\nu)+P(K^0\to\pi^+ e^-\nu)}~.\label{Eq-41}\eeqn
This is equal to $A_T$ in Eq.~\ref{Eq-40} if there is no CPT and no T violation in the $\pi e\nu$ decay amplitudes. In general,
\beq A_T^{exp}(t) = A_T(t) + f(t|{\rm Im}\,x_+,{\rm Re}\,x_-) - 4\,y - 4 {\rm Re}\,x_-~.\label{Eq-42}\eeqn
Assuming CPT symmetry in the decay amplitudes with $y=x_-=0$, CPLEAR presents two results in Ref.~\cite{1998-CPLEAR-T},
the T-violating asymmetry average for $\tau_S<t<20~\tau_S$,
\beq 4~ {\rm Re}~\epsilon + \langle  f(t|{\rm Im}\,x_+)\rangle = (6.6\pm 1.3\pm 1.0)\times 10^{-3}~,\eeqn
and from a two-parameter fit the two T-violating quantities
\beq 4~{\rm Re}~\epsilon = (6.2\pm 1.4\pm 1.0)\times 10^{-3}~{\rm and~Im}\,x_+=(1.2\pm 1.9\pm 0.9)\times 10^{-3}~.\label{Eq-43}\eeqn
Both results violate T symmetry with 4 to 5 $\sigma$, and Re~$\epsilon$ in Eq.~\ref{Eq-43} is in perfect agreement with those from
the earlier Bell-Steinberger analyses.

The time dependences for the survival of $K^0$ ($\Kqz$) states from initial $K^0$ ($\Kqz$) states are obtained 
from Eqs.~\ref{Eq-3} as well; they are found to be
$$   P(K^0\to K^0) = (\frac{1}{4} + {\rm Re}~\delta)\re^{-\Gamma_S t}+(\frac{1}{4}- {\rm Re}~\delta)\re^{-\Gamma_L t}
                     + (\frac{1}{2}\cos\Delta m t -2~{\rm Im}~\delta~\sin\Delta m t)\re^{-\Gamma t} ~,$$
\beq P(\Kqz\to\Kqz) = (\frac{1}{4} - {\rm Re}~\delta)\re^{-\Gamma_S t}+(\frac{1}{4}+ {\rm Re}~\delta)\re^{-\Gamma_L t}
                     + (\frac{1}{2}\cos\Delta m t +2~{\rm Im}~\delta~\sin\Delta m t)\re^{-\Gamma t} ~. \label{Eq-44}\eeqn
The resulting CPT asymmetry
\beq A_{CPT}(t) =\frac {P(\Kqz\to \Kqz)-P(K^0\to K^0)}{P(\Kqz\to \Kqz)-+P(K^0\to K^0)}\label{Eq-45}\eeqn
is time-dependent, sensitive to both Re~$\delta$ and Im~$\delta$, and is equal to 4~Re~$\delta$ for large times, $t\gg\tau_S$.
CPLEAR has determined this asymmetry \cite{1998-CPLEAR-CPT}, again using preparation of the initial state by associate production
in $p\overline p$ annihilation and detection of the final state by a $\pi e\nu$ decay. The experimental CPT asymmetry
\beq A_{CPT}^{exp}(t) =\frac {P(\Kqz\to \pi^+ e^-\nu)-P(K^0\to \pi^- e^+\nu)}{P(\Kqz\to \pi^+ e^-\nu)+P(K^0\to \pi^- e^+\nu)}
\label{Eq-46}\eeqn
is equal to ${\rm 4\,Re\,\delta+2\,y + 2\,Re\,x_-}$ for $t\gg\tau_S$. 
By using their own $\pi^+\pi^-$ decay data, CPLEAR determines a modified
asymmetry $A_\delta$ for eliminating the dependence on $y$ and $x_-$,
\beqa A_{\delta} &=&\frac{N(\Kqz\to\pi^- e^+\nu)-N(K^0\to\pi^+ e^-\nu)(1+4\,{\rm Re}\,\eta_{+-})}
{N(\Kqz\to\pi^- e^+\nu)+N(K^0\to\pi^+ e^-\nu)(1+4\,{\rm Re}\,\eta_{+-})}\nonumber\\ 
            &+& \frac{N(\Kqz\to\pi^+ e^-\nu)-N(K^0\to\pi^- e^+\nu)(1+4\,{\rm Re}\,\eta_{+-})}
{N(\Kqz\to\pi^+ e^-\nu)+N(K^0\to\pi^- e^+\nu)(1+4\,{\rm Re}\,\eta_{+-})}
            = 8\,\rdel+ f(t)~,\eeqan
with ${\rm Re}\,\eta_{+-}=\reps-\rdel$ and $f$ vanishing for $t\gg\tau_S$. For large values of $t$, this quantity 
is strictly independent of all parameters with the exception 
of $\rdel$, \ie~$A_\delta = 2~A_{CPT}$. CPLEAR's result is compatible with zero: CP violation in $K^0\Kqz$ transitions 
is T-violating and not CPT-violating.

\section {$\pi\pi$ Decay Amplitudes} \label{Sec-PiPi}

After many years of effort the two experiments NA48 at CERN \cite{2002-NA48} and KTeV at FNAL \cite{2011-KTeV} have determined final 
results for Re~($\epsilon^\prime/\epsilon_W$); see Eqs.~\ref{Eq-24} and \ref{Eq-25} and Fig.~\ref{Fig-WuYang} for the used notation.
Including a scale factor, the present average \cite{2012-PDG} is
\beq  {\rm Re}\,(\epsilon^\prime/\epsilon_W) = (1.66\pm 0.23)\times 10^{-3}~. \eeqn

\begin{figure}[h!]
\begin{minipage}{0.35\textwidth}
\includegraphics[width=\textwidth]{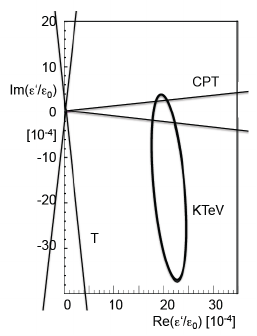}
\caption{$\Delta\chi^2=1$  contour for $\epsilon^\prime/\epsilon_W$ measured by KTeV \cite{2011-KTeV}. 
The two bands marked CPT and T are the allowed regions 
$(\pm 1\,\sigma)$ for CPT symmetry and T symmetry of $\epsilon^\prime$.} \label{Fig-KTeV}
\end{minipage}\hfill\begin{minipage}{0.60\textwidth}
KTeV determines also the phases of $\eta_{+-}$ and $\eta_{00}$ leading to the imaginary part of
$\epsilon^\prime/\epsilon_W$,
\beqa  \phi_{00}-\phi_{+-}&=&(0.30\pm 0.35)^\circ~,\nonumber\\
       {\rm Im}\,(\epsilon^\prime/\epsilon_W) &=& (1.7\pm 2.0)\times 10^{-3}. \eeqan
Since the phase of $\epsilon_W$ is well measured, $\phi(\epsilon_W)=(43.9\pm 0.6)^\circ$ \cite{2011-KTeV}, real and imaginary part of
$\epsilon^\prime/\epsilon_W$ determine the phase of $\epsilon^\prime$. Using the amplitudes
\beqa A_I=\langle \pi\pi,\,I|D|K^0\rangle = a_I\times \re^{\ri\,\delta_I}~,\nonumber\\
      \Aq_I= \langle \pi\pi,\,I|D|\Kqz\rangle = {\overline a}{}_I\times \re^{\ri\,\delta_I}~,\eeqan
where $a_I$ and ${\overline a}{}_I$ are the weak amplitudes and $\delta_I$ the final-state scattering phases 
for isospin $I=0$ and 2, CPT symmetry requires ${\overline a}{}_I=a^*_I$ and \cite{1952-Watson, BrancoLavouraSilva}
\beq     \epsilon^\prime = \frac{1}{\sqrt{2}}\,\frac{p\, a_2-q\, a_2^*}{p\, a_0+q\, a_0^*}\,\frac{\re^{\ri\delta_2}}{\re^{\ri\delta_0}}
                          = \frac{\ri}{\sqrt{2}}\,{\rm Im}\,\frac{a_2}{a_0}\,\re^{\ri(\delta_2-\delta_0)}~.\label{Eq-50}\eeqn
\end{minipage}\end{figure}

{\noindent With $\delta_2-\delta_0=(-45\pm 6)^\circ$ from $\pi\pi$ scattering \cite{1991-GasserMeissner}, we obtain}
\beq  \phi_{CPT}(\epsilon^\prime/\epsilon_W) = \pi/2+\delta_2-\delta_0 + n\,\pi= 45^\circ\pm 6^\circ~{\rm or}~225^\circ\pm6^\circ~.\label{Eq-51}\eeqn
T symmetry requires $(p\, a_2-q\, \aq_2)/(p\, a_0+q\, \aq_0)$ to be real, \ie~$\phi(\epsilon^\prime)=\delta_2-\delta_0+n\pi$,
\beq  \phi_{T}(\epsilon^\prime/\epsilon_W) = -89^\circ\pm 6^\circ~{\rm or}~91^\circ\pm6^\circ~.\label{Eq-52}\eeqn
The two regions for the ratio of the predicted $\epsilon^\prime$ and the measured $\epsilon_W$ are shown in Fig.~\ref{Fig-KTeV}.
The measured $\epsilon^\prime$ is CPT-symmetric and violates T symmetry with about 6 $\sigma$. The observation that the imaginary part
of $a_2/a_0$ is non-zero may be called ``direct T violation".

\section {Summary}

Let me collect here the main results on T violation in the $K^0$-meson system:
\begin{itemize}
\item{CP violation is observed in $K^0_L\to\pi^+\pi^-$ decays (1964). When CP is violated, either CPT or T or both must also be violated.}
\item{The origin is CP violation in $K^0\Kqz$ transitions (1967).}
\item{A unitarity analysis proves that the rate for $\Kqz\to K^0$ is larger than that for $K^0\to \Kqz$, \ie~T violation (1970).}
\item{Unitarity analyses with increased data precision determine the transition parameters\\
      Re\,$\epsilon=(161.1\pm 0.5)\,10^{-5}$, Im\,$\delta=(-0.7\pm 1.4)\,10^{-5}$, Re\,$\delta=(24\pm 23)\,10^{-5}$ (2012).\\ 
      $|m(\Kqz)-m(K^0)| < 6\times 10^{-18}~{\rm or}~< 4\times 10^{-19}$ GeV depending on assumptions.}
\item{T violation in $K^0\Kqz$ transitions is Standard-Model physics owing to three-family quark mixing (GLCKM). This was confirmed
      in 2001 when large CP and T violation were found in B-meson decays. There is no deviation from this conclusion
      with more and more B-meson data until today.}
\item{There is only one more observed T violation in the Kaon system: The complex decay-amplitude ratio $\epsilon^\prime$ violates T
      with about 6 $\sigma$ (2011), T violation in $\pi\pi_{I=2}$ decays.} 
\item{T violation in $\pi\pi_{I=0}$ decays is completely given by Re\,$\epsilon$ owing to unitarity (1970).}
\item{Because of $\Delta Q=\Delta S$, no T violation is seen in semileptonic decay amplitudes (1999).}
\item{No T violation is seen in the transverse muon polarisation of $K\to\pi\mu\nu$ decays (2006).}
\end{itemize}
{\noindent Some more details on the physics of this presentation and more references may be found in a recent review \cite{2014-Schubert}.}
\vspace{8mm}

{\noindent{\bf {\Large Acknowledgements}}}\\[5mm]
I would like to thank Adrian Bevan and his co-organisers of the Conference for the occasion to present this historical review and for
their hospitality. I also thank Hans-J\"urg Gerber and Thomas Ruf for many helpful discussions on T violation
and Philipp Schubert for critically reading the manuscript.

\end{document}